\documentclass[12pt,a4paper,english]{article}
\usepackage[utf8]{inputenc}
\usepackage[english]{babel} 
\usepackage[T1]{fontenc}        
\usepackage{textcomp}

\usepackage{setspace}
\usepackage{amsmath,amsthm,amssymb,amsfonts}
\usepackage{bm}
\usepackage{blkarray}

\usepackage{hyperref}

\usepackage{caption}
\usepackage{subcaption}

\usepackage{natbib}
\usepackage{booktabs, siunitx}
\usepackage{multirow}

\usepackage[margin=2.5cm]{geometry}

\usepackage{pdfpages}
\usepackage{graphicx}

\usepackage[section]{placeins}

\usepackage{tikz}
\usetikzlibrary{fit}
\usetikzlibrary{arrows.meta}
\tikzset{>={Latex[width=2mm,length=2mm]}}

\usepackage{outlines}

\newcommand{\N}{\mathcal{N}}
\newcommand{\R}{\mathbb{R}}
\newcommand{\E}{\mathbb{E}}

\newcommand{\Cov}{\mathrm{Cov}}
\newcommand{\tth}{\footnotesize ^{\mbox{th}}\normalsize}
\newcommand{\deffeq}{\mathrel{\overset{\makebox[0pt]{\mbox{\normalfont\tiny\sffamily def}}}{=}}}
\newcommand{\ie}{{\it i.e., }}

\usepackage{authblk}

\title{A Multivariate Methodology for  Analysing Students' Performance Using Register Data}
\author[1]{Jeanett S. Pelck}
\author[2]{Rafael Pimentel Maia}
\author[2]{Hildete P. Pinheiro}
\author[1,*]{Rodrigo Labouriau}
\affil[1]{Department of Mathematics,  Aarhus University, Denmark}
\affil[2]{Department of Statistics,  University of Campinas, Brazil}
\affil[*]{\footnotesize Corresponding author: Rodrigo Labouriau , rodrigo.labouriau@math.au.dk \normalsize}

\date{February, 2021}

\begin{document}
\maketitle

\begin{abstract}
We present a new method for jointly modelling the students' results in the university's admission exams and their performance in subsequent courses at the university. The case considered involved all the students enrolled at the University of Campinas in 2014 to evening studies programs in educational branches related to exact sciences. 
We collected the number of attempts used for passing the university course of geometry and the results of the admission exams of those students in seven disciplines. The method introduced involved a combination of multivariate generalised linear mixed models (GLMM) and graphical models for representing the covariance structure of the random components. The models we used allowed us to discuss the association of quantities of very different nature. 
We used Gaussian GLMM for modelling the performance in the admission exams and a frailty discrete-time Cox proportional model, represented by a GLMM, to describe the number of attempts for passing Geometry.
 
The analyses were stratified into two populations: the students who received a bonus giving advantages in the university's admission process to compensate social and racial inequalities and those who did not receive the compensation. The two populations presented different patterns. Using general properties of graphical models, we argue that, on the one hand, the predicted performance in the admission exam of Mathematics could solely be used as a predictor of the performance in geometry for the students who received the bonus. On the other hand, the Portuguese admission exam's predicted performance could be used as a single predictor of the performance in geometry for the students who did not receive the bonus.
\end{abstract}

\normalsize

\newpage



\section{Introduction}
In this paper, we study the admission system to a Brazilian university and the bonus system for compensating social and racial inequalities. The data analysed below is based on the registers of entrance and performance at the University of Campinas, Brazil (UNICAMP).
In 2005, UNICAMP implemented an affirmative action program giving extra bonus in the final entrance examination score for students who were enrolled for their entire high school years in the public system (with an additional bonus for those who self declared to be African / Indigenous Brazilian descendants). 
See \cite{Maia2016,Pedrosa2007,Pinheiro2019,Pinheiro2020} for more details.

The main difficulty of the study of those registers is the multivariate nature of the characterisations of the object of interest, and the (unavoidable) presence of spurious associations. The responses observed in the data referred above are of very different nature but can be analysed in one multivariate model as we will describe below. The performance at the university is measured by the number of attempts required to pass the course of geometry, which is a key course in the beginning of the university education of the group of students we study.
This response is typically right-censored, in the sense that there might be some students that have not passed the course when the data was collected, dropped out during the study or who's enrolment has been cancelled. 
The enrolment is cancelled if the student fails all the subjects in the first or second semester or reached the maximum number of semesters allowed without graduating (e.g. in Statistics this is 6 years). 
On the other hand, the performances at the admission exams are measured in a standardised scale using a scoring system.
Those two types of responses are modelled differently: the time for passing a course is modelled using a variant of the frailty Cox proportional model with discrete-time; the scores at the entrance exam are described using a Gaussian mixed model. 
In both cases, the models can be represented as instances of generalised linear mixed models (GLMMs). The introduced GLMMs contain two common random components (one taking  different values for each individual, and one taking the same value for individuals enrolled at the same branch of study). The random component related to the individuals allow us to connect the models describing the different responses, and in this way, characterise how much information each response carries on the other responses. This methodology differs from \cite{Pinheiro2020} where parts of the same data were analysed but in a different context. 

The analyses performed were stratified into two populations: the students who received a bonus, and the students who did not receive the compensation. The two populations are different and presented different patterns, justifying the stratification. 

This study aims to present statistical tools that allow to study the different facets of the type of data described above and to understand the associations between the different responses. 
These aims are fulfilled by using suitable multivariate versions of GLMMs and by using the theory of graphical models to describe the covariance structure of the common random component giving the variation between individuals \citep{Pelck2021B}.
We illustrate, in this way, a process of modelling responses of very different nature in a multivariate model that arises when working with educational register data. 

The paper is organised as follows. Section \ref{Section:Data} describes the data used. 
The multivariate GLMM are introduced in Section \ref{sec:modelDes}, including details on the marginal Gaussian GLMMs and the frailty discrete-time Cox Proportional model.
Section \ref{sec:GraphMod} describes the graphical model used for representing the covariance structure of the random components, and Section \ref{sec:Results} presents and discusses the results. Appendix \ref{SectionAppendixMC} presents some model control, while some details of the representation of the graphical models are given in appendix \ref{SectionAppendixGR}.

\section{Data Description}
\label{Section:Data}
The data we used contain records on all the $299$ students enrolled at the UNICAMP in 2014, in evening studies programs in one of the educational branches related to exact sciences listed in Table \ref{TAB01}. 
Among those students, $151$ received a bonus giving advantages in the university's admission process. The bonus group consists of students from a public high school and students from a public high school who are self-declared African or Indigenous Brazilian descendants.

\begin{table}[ht]
\begin{tabular}{lll}
\toprule
Chemical engineering &
Electrical engineering &
Economical science \\
Mathematics &
Physics &
Computer science \\
Automation engineering &
Technological chemistry &
Bachelor in Chemistry \\
 Medical Physics & &
 \\
  \bottomrule
 \end{tabular}
\caption{Educational branches included in the population studied.}
\label{TAB01}
\end{table}

\noindent
The data includes eight responses recorded for each student. The first seven responses correspond to the student's performance in the university admission exam in the disciplines: 
Mathematics, Physics, Chemistry, Biology, History, Geography, and Portuguese. The last response was the number of attempts the students used to pass the (first year) geometry course at the university. This response was right censored (with $24.08\%$ of censoring) since some students used at least the observed number of attempts, but it was unknown whether the enrolment had been cancelled or the student passed the course after the data was collected. Additionally, the registers contain a range of information on each individual including gender and age.

\section{A Multivariate Model for for Simultaneously Describing the Admission Scores and the Performance in Geometry\label{sec:modelDes}}

 The eight responses described above were jointly analysed using a multivariate generalised linear mixed model as described below. We performed separate parallel analyses for the students who received the bonus and those who did not receive the bonus, which, we anticipate, will yield contrasting results.
 
 In each of the two separate analyses, we used a multivariate model combining seven Gaussian mixed models describing the seven admission exams, and a frailty Cox proportional model with discrete-time for modelling the number of attempts to pass the course of geometry. Each of the marginal models above included two random components: one accounting for the variation between the different study branches, and one representing each individual. The eight marginal models referred above were combined by assuming a joint multivariate Gaussian distribution for the random component representing the individuals. The precise model definition for the students who received bonus is given below. The models for the students that did not receive bonus are similarly defined. 
 
 In order to describe the models we will use, we index the individuals by $i$ ($i = 1, \ldots, n$, with $n=151$), the eight responses by $j$ ($j=1,\dots ,8$) and the educational branches listed in Table \ref{TAB01} by $k$ ($k=1,\ldots,10$). 
 Moreover, the educational branch of the $i\tth$ student is denoted by $e(i)$. We describe below the covariance structure of the multivariate generalised linear mixed models we want to introduce using two random components.  
 The random component representing the educational branches is defined by assuming that there exist $10$ unobservable random variables $U_1^{[j]},\ldots, \, U_{10}^{[j]}$ for each of the eight responses ($j=1,\ldots,8$) corresponding to the $10$ educational branches. The random variable $U_k^{[j]}$ takes the same value for the $j\tth$ response for each student that is enrolled in the $k\tth$ educational branch ($k=1, \ldots, 10$, $j=1, \ldots, 8$). 
 The random components representing the individuals is specified for the $j\tth$ response ($j=1,\ldots,8$) by defining the unobservable random variables $V_1^{[j]},\ldots,V_{n}^{[j]}$ representing the $n$ individuals.  

 According to the model, for $j, j' = 1,\ldots, 8$, the random vectors 
%
${\bm{U}}^{[j]} \deffeq  (U_1^{[j]},\ldots,U_{10}^{[j]} )^T$ and 
${\bm{V}}^{[j']} \deffeq  (V_1^{[j']},\ldots,V_{n}^{[j']} )^T$
 are independent and multivariate Gaussian distributed as given below, 
 \begin{align*}
	{\bm{U}}^{[j]} \sim N_{10} \left (\bm{0},\sigma^2_{{U}^{[j]}}\bm{I}_{10}\right )\\
	{\bm{V}}^{[j']} \sim N_{n} \left (\bm{0},\sigma^2_{{V}^{[j']}}\bm{I}_{n}\right ).
\end{align*}
 Here $\bm{I}_{m}$ denotes a $m$-dimensional identity matrix (for $m \in \N$).
 Furthermore, for $j,j' = 1,\ldots ,8$, with $j\neq j'$, we assume that $\bm{U}^{[j]}$ is independent of $\bm{U}^{[j']}$,  and  that  $V_i^{[j]}$ is independent $V_{i'}^{[j']}$, where $i, i' = 1, \ldots , n$ with $i\ne i'$ .
 Additionally, we assume that for $i=1,\ldots,n$,
\begin{align} \label{CovVs}
\Cov \left ( V_i^{[1]},\ldots,V_i^{[8]} \right )=\bm{\Sigma}_V,
\end{align}
with $\text{diag} \left (\bm{\Sigma}_V\right ) = \left (\sigma_{V^{[1]}}^2, \ldots, \sigma_{V^{[8]}}^2\right )$.
Therefore, defining $\bm{V}^T \deffeq \left ( {\bm{V}^{[1]}}^T , \ldots ,  {\bm{V}^{[8]}}^T \right )$ we have that
\begin{align*}
\Cov \left ( \bm{V} \right ) =  \bm{\Sigma}_{V} \otimes \bm{I}_{n} \, . 
\end{align*}
We will use the matrix $\bm{\Sigma}_V$ to characterise the dependence structure of the eight responses. In particular, in Section~\ref{sec:GraphMod}, we will use a graphical model structure by imposing zeroes in the inverse of $\bm{\Sigma}_V$, corresponding to assume that some pairs of the random variable ${{V_i}}^{[1]}, \ldots , {{V_i}}^{[8]}$ are conditionally independent given the other random variables.
For improving the readability of the discussion below, when relevant, we denote  ${{V_i}}^{[1]}, \ldots , {{V_i}}^{[8]}$ by 
${{V_i}}^{[Math]}, \ldots , {{V_i}}^{[Geom]}$,\ie we identify the superindices of the individual random components with a recognisable short form of the corresponding response. 

We formulate the marginal generalised linear mixed models for the seven responses related to the admission exams by specifying the conditional expectations for the $i$th individual ($i=1,\ldots,n$) in the conditional Gaussian distributions with identity link function given $(U^{[j]}_{e(i)},V_i^{[j]})$ for $j=1,\ldots,7$, that is, 
\begin{align*}
	\E [Y_i^{[j]}\vert U^{[j]}_{e(i)}=u, \, V_i^{[j]}=v] = \bm{x}_i^T \bm{\beta}^{[j]}+u+ v, \quad \forall \, u,v\in\R.
\end{align*}
In all the models above, the term
$\bm{x}_i^T \bm{\beta}^{[j]}$ ($j=1,\ldots,7$) adjusts for gender (female or male) and age with age divided into two groups (under 21 or 21 and above).

We formulate the discrete time frailty Cox proportional hazard model describing the number of attempts to pass the course of geometry.
According to the model, for the student $i\tth$ ($i = 1, \ldots, n$), the discrete conditional  hazard function given $(U^{[8]}_{e(i)},V_i^{[8]})$ is
\begin{eqnarray} \nonumber
\lambda_i(t \vert U^{[8]}_{e(i)}=u, \, V_i^{[8]}=v) & \deffeq &
P( T_i = t  \vert T_i \ge t, U^{[8]}_{e(i)}=u, \, V_i^{[8]}=v) \\ \nonumber
 & = &
\tilde\lambda_{t}\exp(\bm{x}_i^T \bm{\beta}^{[8]}) \exp(u) \exp(v), \mbox{ for }t = 1,2, \ldots \mbox{ for all } u,v\in \R.
\end{eqnarray}
Here $T_i$ is the random variable representing the number of attempts to pass the course of geometry used by the $i\tth$ individual and the term
$\bm{x}_i^T \bm{\beta}^{[8]}$ adjusts for gender (female or male) and age with age divided into two groups (under 21 or 21 and above).
The model above coincides with a generalised linear mixed model defined with a binomial distribution and a logarithmic link function, applied to a specially constructed data representing the risk set of the related counting process. 
We used a Poisson approximation for avoiding numerical issues.
See \cite{Maia2014}. 

\section{Modelling the Covariance Structure of the Random Components \label{sec:GraphMod}} 

We complete the specification of the multivariate generalised linear mixed model introduced in Section \ref{sec:modelDes} by defining the covariance structure of the random components representing the individual's variation. 
Since the random components ${V}_i^{[1]}, \ldots, {V}_i^{[8]}$ have the same distribution for $i=1, \ldots, n$, we suppress the subindex $i$ from the notation and write ${V}^{[1]},\ldots,{V}^{[8]}$ to denote ${V}_i^{[1]},\ldots,{V}_i^{[8]}$ for an arbitrary individual.

The random components   ${V}^{[1]},\ldots,{V}^{[8]}$  represent the individual variation of the abilities of each of the $n$ students affecting the performance in the seven admission exams and in the course of geometry, respectively. Note that according to the model, the covariance of the random variables $V_i^{[1]}, \ldots, V_i^{[8]}$ is the same for all the individuals, namely $\bm{\Sigma}_V$, see (\ref{CovVs}). 
Here we will characterise this covariance structure common to all the individuals using graphical models, which will allow us to draw general conclusions on the interdependence between the eight responses studied. Before pursuing this task, we give a short account of the theory of graphical models; for a comprehensive description of this theory see \cite{Lauritzen1996} and \cite{Whittaker1990}.

Let $\mathcal{G}=(\mathcal{V},\mathcal{E})$ denote a graph with a set of vertices, $\mathcal{V}$, and edges, $\mathcal{E}\subseteq \mathcal{V}\times \mathcal{V}$.
Each vertex represents a random variable, and two vertices are connected with an edge if, and only if, they are not conditionally independent given the remaining random variables. We say that there is a path between two vertices if there exist a sequence of pairs of vertices connected with an edge connecting the two vertices. 

In the multivariate models described above, we consider a graph, $\mathcal{G}=(\mathcal{V},\mathcal{E})$, with $\mathcal{V}= \{ {V}^{[1]},\ldots,{V}^{[8]} \} =  \{ {V}^{[Math]},\ldots,{V}^{[Geom]} \}$.
The set of edges contains pairs of random variables which are not conditionally independent given the remaining random variables, and thus, they carry some information on each other that are not contained in the other random variables in $\mathcal{V}$. For example, suppose that there is an edge between 
${V}^{[Geom]}$ and ${V}^{[Math]}$ associated to the individual responses related to the performance in the course of geometry and the admission exam in mathematics, respectively. This means that after correcting for differences in age, gender and education branch, the random variables ${V}^{[Geom]}$ and  $ {V}^{[Math]}$ are conditionally dependent given $ \mathcal{V} \setminus \left \{{V}^{[Geom]}, {V}^{[Math]}\right \}$; therefore ${V}^{[Math]}$ carries information on ${V}^{[Geom]}$ that is not contained in the informational contents of the random variables  $ \mathcal{V} \setminus \left \{{V}^{[Geom]}, {V}^{[Math]}\right \}$. Conversely, the absence of an edge connecting two vertices indicates that the random variables associated to those vertices are conditionally independent given the other random variables in play; therefore, the knowledge of the other random variables renders the two random variables in question independent.

According to the theory of graphical models, a set  of vertices, say $\mathcal{S}$, separates two sets of vertices $\mathcal{A}$ and $\mathcal{B}$ in the graph, if, and only if,  each path connecting an element of $\mathcal{A}$ to an element of $\mathcal{B}$ contains at least one element of $\mathcal{S}$. A key result in the theory of graphical models is that if a set of vertices, $\mathcal{S}$, separates two disjoint subsets of vertices $\mathcal{A}$ and $\mathcal{B}$, then all the variables in $\mathcal{A}$ are independent of the variables in $\mathcal{B}$ given the variables in $\mathcal{S}$. This result is called the \emph{separation principle} or \emph{global Markov property} for undirected graphical models \cite[page 32]{Lauritzen1996}. 
For example, if the random variable ${V}^{[Math]}$ separates the random variable ${V}^{[Geom]}$ from the other vertices, then conditioning on solely  ${V}^{[Math]}$ renders ${V}^{[Geom]}$ independent of the other vertices (\ie $ \mathcal{V} \setminus \left \{{V}^{[Geom]}, {V}^{[Math]}\right \}$). 

Note that the graph $\mathcal{G}=(\mathcal{V},\mathcal{E})$ defined with $\mathcal{V}= \{ {V}^{[Math]},\ldots,{V}^{[Geom]} \}$ involves the unobservable random variables associated with the individual random components, not the observed responses. However, it is possible to extend the separation principle using the multivariate generalised linear mixed model's properties to discuss the interdependence of the observed responses. We explain this extended principle using an example. Suppose that the random variable $ {V}^{[Math]}$  separates the random variable ${V}^{[Geom]}$ from the other vertices in $\mathcal{V}$, then, according to the extended separation principle, each 
response corresponding to a vertex in  $ \mathcal{V} \setminus \left \{{V}^{[Geom]}, {V}^{[Math]}\right \}$ , say $Y_i^{[*]}$, is independent of  $T_i = T_i^{[Geom]}$ (\ie the number of attempts  the $i\tth$ individual uses to pass the course of geometry)
 given ${V}^{[Math]}$.
Here it is required to condition on the random variable ${V}^{[Math]}$ for obtaining independency of  $Y_i^{[*]}$  and $T_i$; conditioning on the observable responses $Y_i^{[Math]}$ does not necessarily renders the variables $Y_i^{[*]}$  and $T_i$ independent.  
See the Appendix \ref{SectionAppendixGR}and \cite{Pelck2021B} for a general formulation of the extended separation principle using undirected graphical models. 
 

The graph $\mathcal{G}=(\mathcal{V},\mathcal{E})$ with vertices $\mathcal{V}=\{V^{[1]}, \ldots, V^{[8]}\}$ was inferred using predicted values of the random variables to infer a graphical model that minimises the BIC. A method and implementation for minimising the BIC are described in \cite{Abreu2010}, and an inference method for predicting values of the random components is presented in \cite{Pelck2021A}. Notice, that this method only yields normally distributed predictors in cases with small variance of the random components. However, in  \cite{Labouriau1998} it is shown that treating graphical models involving non Gaussian random variables as being normally distributed corresponds to using optimal inferential procedure for semiparametric models under mild regularity conditions. 

The survival model presented was controlled using the methods described in  \cite{Maia2014,Edwards2010}. We found no indication of a lack of fit of the models. The marginal Gaussian mixed models where controlled by standard residual analyses. See appendix \ref{SectionAppendixMC}. 

\section{Results and Discussion}
\label{sec:Results}

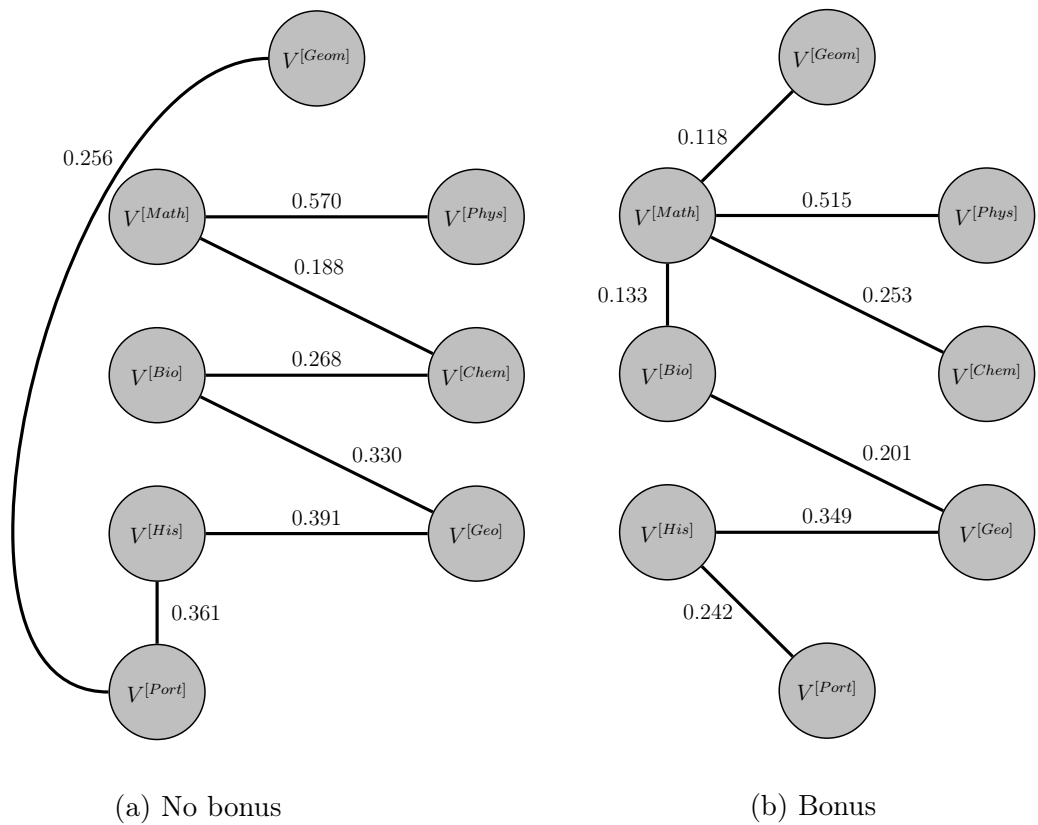
\begin{figure}
\begin{subfigure}{0.5\textwidth}
\begin{center}\scalebox{0.7}{
 \begin{tikzpicture}[
                                                                                                                                                 squarednode/.style={rectangle, draw=black, fill=lightgray, thick},
                                                                                                                                                 roundnode/.style={circle, draw=black, fill=lightgray, thick},
                                                                                                                                                 ]
                                                                                                                                                 \node [style=roundnode,minimum size=1.8cm] (His) at (-3, 3) {$V^{[His]}$};
                                                                                                                                                 \node [style=roundnode,minimum size=1.8cm] (Chem) at (3, 6) {$V^{[Chem]}$};
                                                                                                                                                 \node [style=roundnode,minimum size=1.8cm] (Port) at (-3, 0) {$V^{[Port]}$};
                                                                                                                                                 \node [style=roundnode,minimum size=1.8cm] (Geo) at (3, 3) {$V^{[Geo]}$};
                                                                                                                                                 \node [style=roundnode,minimum size=1.8cm] (Math) at (-3,9) {$V^{[Math]}$};
                                                                                                                                                 \node [style=roundnode,minimum size=1.8cm] (Bio) at (-3, 6) {$V^{[Bio]}$};
                                                                                                                                                 \node [style=roundnode,minimum size=1.8cm] (Phys) at (3, 9) {$V^{[Phys]}$};
                                                                                                                                                 \node [style=roundnode,minimum size=1.8cm] (Geom) at (0, 12) {$V^{[Geom]}$};
                                                                                                                                        \draw [-,line width=1.7] (His) to node[midway,above=0cm,align=center]{0.391} (Geo);
                                                                                                                                                 \draw [-,line width=1.7] (His) to node[midway,right=0.1cm,align=center]{0.361}(Port);
                                                                                                                                                 \draw [out =180,in=180,looseness=0.8,line width=1.7] (Geom) to node[near start,left=0.1cm,align=center]{0.256}(Port);
                                                                                                                                                 \draw [-,line width=1.7] (Phys) to node[midway,above=0cm,align=center]{0.570}(Math);
                                                                                                                                                 \draw [-,line width=1.7] (Chem) to node[near end,right=0.5cm,align=center]{0.188}(Math);
                                                                                                                                                 \draw [-,line width=1.7] (Bio) to node[midway,above=0cm,align=center]{0.268} (Chem);
                                                                                                                                                 \draw [-,line width=1.7] (Bio) to  node[midway,right=0.5cm,align=center]{0.330} (Geo);
                                                                                                                                                 
 \end{tikzpicture}
 }
 \end{center}
 \caption{No bonus}
 \end{subfigure}
 \begin{subfigure}{0.5\textwidth}
\begin{center}\scalebox{0.7}{
 \begin{tikzpicture}[
                                                                                                                                                 squarednode/.style={rectangle, draw=black, fill=lightgray, thick},
                                                                                                                                                 roundnode/.style={circle, draw=black, fill=lightgray, thick},
                                                                                                                                                 ]
                                                                                                                                                 \node [style=roundnode,minimum size=1.8cm] (His) at (-3, 3) {$V^{[His]}$};
                                                                                                                                                 \node [style=roundnode,minimum size=1.8cm] (Chem) at (3, 6) {$V^{[Chem]}$};
                                                                                                                                                 \node [style=roundnode,minimum size=1.8cm] (Port) at (0, 0) {$V^{[Port]}$};
                                                                                                                                                 \node [style=roundnode,minimum size=1.8cm] (Geo) at (3, 3) {$V^{[Geo]}$};
                                                                                                                                                 \node [style=roundnode,minimum size=1.8cm] (Math) at (-3,9) {$V^{[Math]}$};
                                                                                                                                                 \node [style=roundnode,minimum size=1.8cm] (Bio) at (-3, 6) {$V^{[Bio]}$};
                                                                                                                                                 \node [style=roundnode,minimum size=1.8cm] (Phys) at (3, 9) {$V^{[Phys]}$};
                                                                                                                                                 \node [style=roundnode,minimum size=1.8cm] (Geom) at (0, 12) {$V^{[Geom]}$};
                                                                                                                                                                                                                                                                                                \draw [-,line width=1.7] (His) to node[midway,above=0cm,align=center]{0.349} (Geo);
                                                                                                                                                 \draw [-,line width=1.7] (His) to node[midway,left=0.1cm,align=center]{0.242}(Port);
                                                                                                                   \draw [-,line width=1.7] (Geom) to node[midway,left=0.2cm,align=center]{0.118}(Math);
                                                                                                                                                 \draw [-,line width=1.7] (Phys) to node[midway,above=0cm,align=center]{0.515}(Math);
                                                                                                                                                 \draw [-,line width=1.7] (Bio) to node[midway,left=0.2cm,align=center]{0.133}(Math);
                                                                                                                                                 \draw [-,line width=1.7] (Chem) to node[midway,right=0.5cm,align=center]{0.253}(Math);
                                                                                                                                                 \draw [-,line width=1.7] (Bio) to  node[midway,right=0.5cm,align=center]{0.201} (Geo);
                                                                                                                                                 
\end{tikzpicture}
}
 \end{center}
 \caption{Bonus}
 \end{subfigure}
\caption{Independence graph representing the estimated graphical model describing the covariance structure of the individual random components $V_i^{[1]},\ldots,V_i^{[8]}$ for an  arbitrarily chosen individual  ($i=1, \ldots, n$, suppressing the index $i$ in the graph). 
The estimated conditional correlations are reported for each edge .
}
\label{figGRAPH}
\end{figure}

Figure \ref{figGRAPH} displays the graphs representing the estimated graphical models describing the covariance structure of the individual random components for the students that received bonus and the students that did not. The two populations of students presented different covariance structures which we discuss below. We stress that each of the individuals' random components represents latent individual abilities affecting the performance related to the respective  responses. The covariance structures described in Figure \ref{figGRAPH} are obtained after adjusting for differences in age, gender and educational branch.

In the population of students that received bonus, the random component related to the performance in the course of geometry, ${V}^{[Geom]}$, is only connected to the random component related to the result in the admission exam of mathematics, ${V}^{[Math]}$ (see the left panel in Figure \ref{figGRAPH} and Figure \ref{fig:AppBonus}). The conditional correlation between ${V}^{[Geom]}$ and ${V}^{[Math]}$ is positive.
This result suggests the existence of common cognitive mechanisms associated with the latent abilities related to the performance in the admission exam of mathematics and in the course of geometry. Since ${V}^{[Math]}$ separates ${V}^{[Geom]}$ from the other individual random components, according to the separation principle, ${V}^{[Geom]}$ is conditionally independent of the other individual random components given ${V}^{[Math]}$. 
This conditional independence indicates that the putative common cognitive mechanisms 
referred above are specific to these two disciplines and are not shared by the other disciplines' abilities.
Regarding the results of the admission exams, according to the extended separation principle, the observed performance in the course of geometry is conditionally independent of the observed results of the admission exams given the individual random component ${V}^{[Math]}$.
From the practical point of view, the result shows that the prediction of the random component ${V}^{[Math]}$ suffices for predicting the performance of the students that received the bonus in the course of geometry. After predicting ${V}^{[Math]}$, both the results of the other admission exams and their corresponding individual random components become uninformative concerning the performance in the course of geometry.

We obtain a different scenario for the population of the students that did not receive the bonus. There, the individual random component ${V}^{[Port]}$ (associated with the performance in the admission exam of Portuguese) is the only random component connected with the random component ${V}^{[Geom]}$; moreover, ${V}^{[Port]}$ separates ${V}^{[Geom]}$ from the other individual random components (see the right panel of Figure \ref{figGRAPH} and Figure \ref {fig:AppNobonus}). Therefore, using a similar rationale as above, we conclude that for the population of students that did not receive the bonus, the prediction of  ${V}^{[Port]}$ suffices for predicting the performance of the students that received the bonus in the course of geometry. 

The variance of the predictions of  ${V}^{[Math]}$ is $43\%$ larger in the population of students that received the bonus, as compared with the variance in the group that did not receive the bonus. A combination of two factors might cause this difference: in the population of students who received the bonus, there might be more considerable variability in the quality of the high school teaching in Mathematics; furthermore, the students with a lower level in Mathematics could enter the University by receiving a bonus. Therefore, the mathematics skills detected in the admission exam play an essential role in the performance in geometry among the students who received the compensation. One might speculate whether the random component ${V}^{[Port]}$  is representing social-economic class which plays a key role in the performance in the course of geometry population of students that did not receive the bonus. 

There are also some similarities between the inferred covariance structures of the two populations of students studied. For example, in both populations, the individuals random component ${V}^{[Bio]}$ (related to the performance in the admission exam of biology) separates ${V}^{[Math]}$, ${V}^{[Phys]}$ and ${V}^{[Chem]}$ from ${V}^{[His]}$ and ${V}^{[Geo]}$. We let the reader explore further aspects of the results presented here. In this paper, we have exposed a new method based on a combination of multivariate generalised linear mixed models and graphical models for modelling and predicting students' performance using responses of different nature, namely, some Gaussian responses and a discrete right-censored response. Other response types can also be modelled by choosing different distributions and different link functions for constructing the marginal models.

\section*{Acknowlegment}

We acknowledge the admission committee (CONVEST) from the University of Campinas (UNICAMP) for giving access to the data used. The first and the last authors were partially financed by the  Applied Statistics Laboratory (aStatLab) from the Department of Mathematics, Aarhus University. The third author was partially financed by Conselho Nacional de Desenvolvimento Cient{\' \i}fico e Tecnol{\' o}gico (CNPq), Grant number: 310874/2018-1.

\bibliography{bibtexFile}

\appendix
 \section{Some Model Control}\label{SectionAppendixMC}
 
\begin{figure}
\begin{subfigure}{0.51\textwidth}
\begin{center}\scalebox{1.7}{
  \includegraphics[width=.55\linewidth]{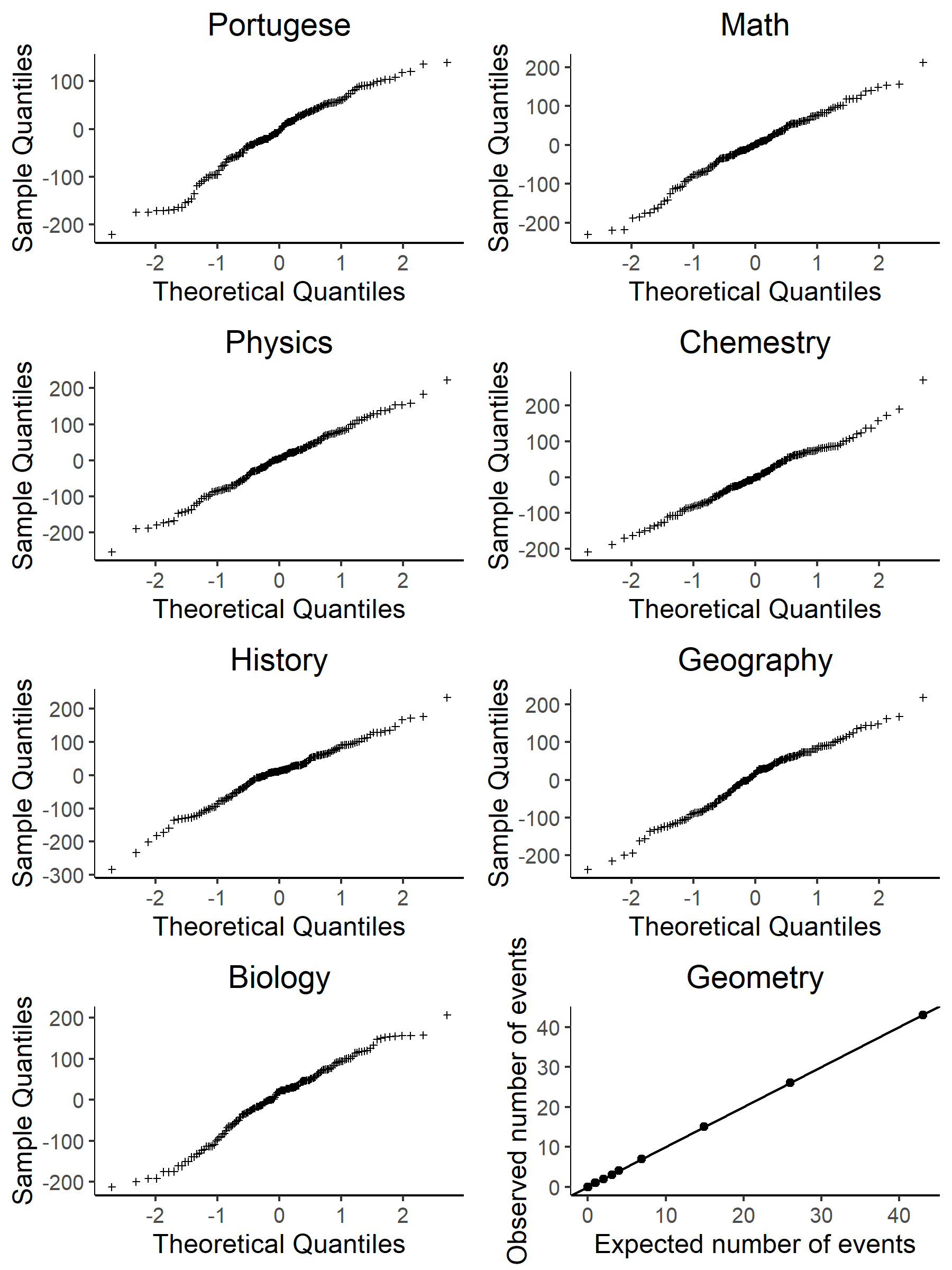}
 }
 \end{center}
 \caption{No bonus}
 \end{subfigure}
 \begin{subfigure}{0.51\textwidth}
\begin{center}\scalebox{1.7}{
  \includegraphics[width=.55\linewidth]{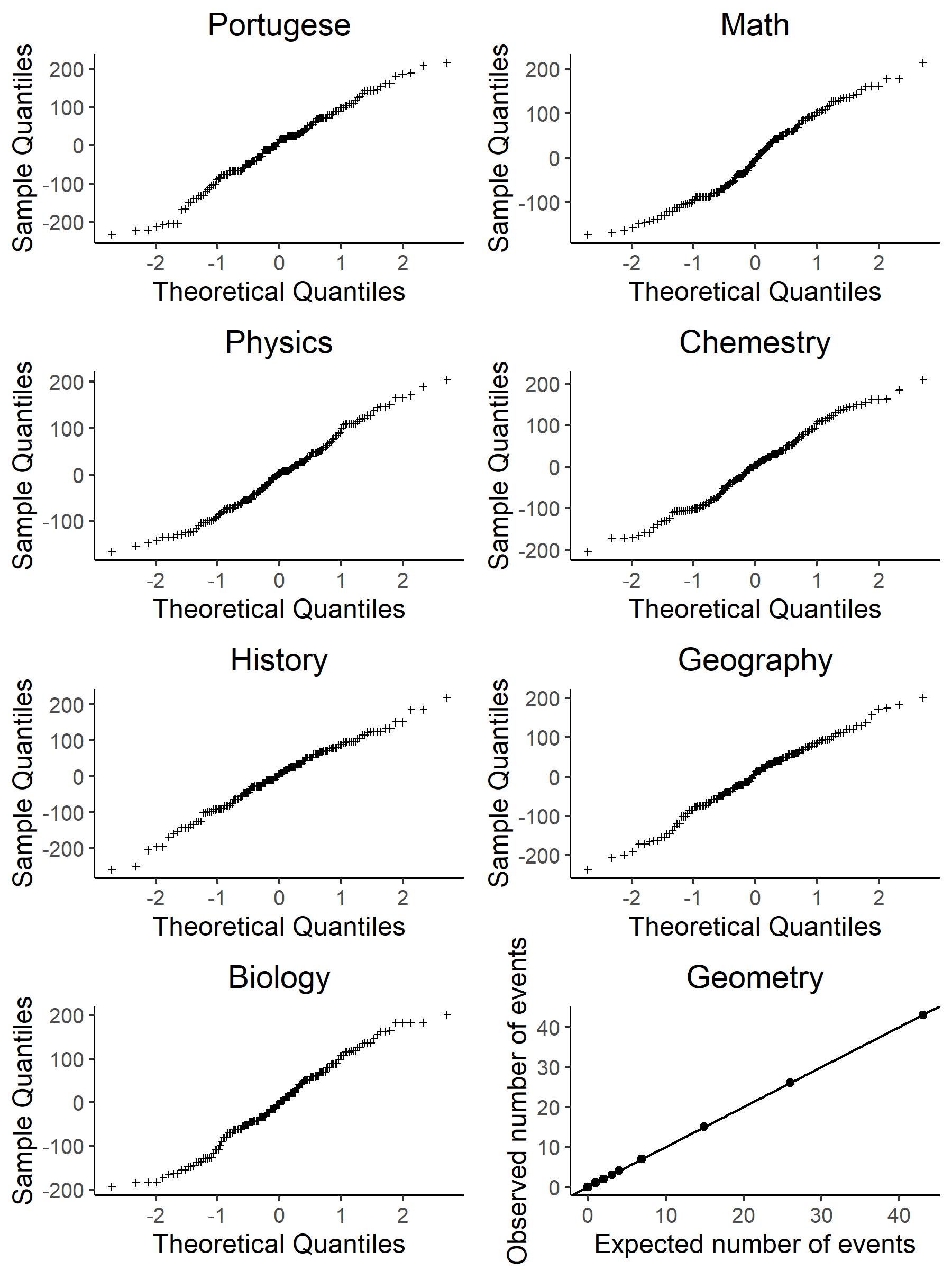}
 }
 \end{center}
 \caption{Bonus}
 \end{subfigure}
\caption{Normal QQ-plot of the responses related to the seven admission exams and a scatter plot of the observed number of events versus expected number of events for each time and combination of age and gender.
\label{fig:MC} }
\end{figure}

 We briefly discuss below the validity of the models used.  
 The residual analyses in the marginal Gaussian mixed models, representing the the responses related to the seven admission exams, show that there is no indication of serious lack of fit, see Figure~\ref{fig:MC}.

Comparing the observed and the expected number of students that passed the course of geometry at each time for each combination of age and gender allowed us to conclude that there is no evidence of lack of global adjustment of the survival model. More precisely,
the predicted number of events at time $t$ ($t=1,\ldots,T$), denoted $n(t)$, is calculated as the number of individuals "at risk" (the number of students that have not passed the course yet and are still studying the course) times the average estimated hazard at time $t$.  More precisely,
\begin{align*}
               n(t)=\vert R_t\vert \sum_{i \in R_t}\hat{\lambda}_{t}\exp(\bm{x}_i^T \hat{\bm{\beta}}) \exp(\hat{u}_{e(i)}) \exp(\hat{v}_{i}),
\end{align*}
where $R_t=\{i \in \{1,\ldots,n\}: t_i\leq t\}$ denote the set of all individuals at risk at time $t$ and $\vert R_t\vert$ the number of individuals in $ R_t$.
The results can be found in Figure~\ref{fig:MC}.

 \newpage
\section{Detailed Representation of the Graphical Models Involving the Random Components and the Response variables}
\label{SectionAppendixGR}

For the reader acquaint with the theory of graphical models (see \citeauthor{Whittaker1990}, \citeyear{Whittaker1990}), the extended separation principle can be formulated in general by defining an directed acyclic graph (DAG, \ie a graph formed by vertices and directed edges represented by arrows obtained by eliminating the symmetry property in the set of edges $\mathcal{E}$). Using basic properties of the generalised linear mixed models of the type discussed here and the factorisation of the joint densities of the distributions of the individual random component, it is possible to show that the interdependence of the the observable responses and the random components related to the individuals can be represented by an acyclic graphs, where there is an arrow from each random component pointing to the random variables representing the  corresponding observable responses. Additionally, the graphical representation referred above contains an undirected edge 
connecting the vertices that are not conditionally independent in the graph representing 
the individual random components (see  \citeauthor{Pelck2021B}, \citeyear{Pelck2021B} for the detailed construction).
Noting that this acyclic graph satisfies the Wermuth condition (see \citeauthor{Whittaker1990}, \citeyear{Whittaker1990}, page 75), which implies that the moral graph obtained, in this case, by making all the edges undirected, satisfies the separation principle (see \citeauthor{Whittaker1990}, \citeyear{Whittaker1990},  theorem 3.5.2 on page 76). This construction yields the extended separation principle which states that {\it "two sets of observable responses, say $\mathcal{\tilde A}$  and $\mathcal{\tilde B}$, are conditionally independent given a set $\mathcal{S}$ of  individual random components, provided $\mathcal{S}$ separates the sets of individual random components $\mathcal{A}$ and $\mathcal{B}$ corresponding to the sets of observable responses $\mathcal{\tilde A}$ and $\mathcal{\tilde B}$}".

\begin{figure}
\begin{center}\scalebox{0.9}{

 \begin{tikzpicture}[
                                                                                                                                                 squarednode/.style={rectangle, draw=black, fill=lightgray, thick},
                                                                                                                                                 roundnode/.style={circle, draw=black, fill=lightgray, thick},
                                                                                                                                                 ]
                                                                                                                                                 \node [style=roundnode,minimum size=1.8cm] (His) at (-3, 3) {$V^{[His]}$};
                                                                                                                                                 \node [style=roundnode,minimum size=1.8cm] (Chem) at (3, 6) {$V^{[Chem]}$};
                                                                                                                                                 \node [style=roundnode,minimum size=1.8cm] (Port) at (-3, 0) {$V^{[Port]}$};
                                                                                                                                                 \node [style=roundnode,minimum size=1.8cm] (Geo) at (3, 3) {$V^{[Geo]}$};
                                                                                                                                                 \node [style=roundnode,minimum size=1.8cm] (Math) at (-3,9) {$V^{[Math]}$};
                                                                                                                                                 \node [style=roundnode,minimum size=1.8cm] (Bio) at (-3, 6) {$V^{[Bio]}$};
                                                                                                                                                 \node [style=roundnode,minimum size=1.8cm] (Phys) at (3, 9) {$V^{[Phys]}$};
                                                                                                                                                 \node [style=roundnode,minimum size=1.8cm] (Geom) at (0, 12) {$V^{[Geom]}$};
                                                                                                                                                 \node [style=squarednode,minimum size=1.5cm] (RHis) at (-8, 3) {$Y^{[His]}$};
                                                                                                                                                 \node [style=squarednode,minimum size=1.5cm] (RChem) at (8, 6) {$Y^{[Chem]}$};
                                                                                                                                                 \node [style=squarednode,minimum size=1.5cm] (RPort) at (-8, 0) {$Y^{[Port]}$};
                                                                                                                                                 \node [style=squarednode,minimum size=1.5cm] (RGeo) at (8, 3) {$Y^{[Geo]}$};
                                                                                                                                                 \node [style=squarednode,minimum size=1.5cm] (RMath) at (-8,9) {$Y^{[Math]}$};
                                                                                                                                                 \node [style=squarednode,minimum size=1.5cm] (RBio) at (-8, 6) {$Y^{[Bio]}$};
                                                                                                                                                 \node [style=squarednode,minimum size=1.5cm] (RPhys) at (8, 9) {$Y^{[Phys]}$};
                                                                                                                                                 \node [style=squarednode,minimum size=1.5cm] (RGeom) at (8, 12) {$T^{[Geom]}$};
                                                                                                                                                 \draw [-,line width=1.7] (His) to node[midway,above=0cm,align=center]{} (Geo);
                                                                                                                                                 \draw [-,line width=1.7] (His) to node[midway,right=0.1cm,align=center]{}(Port);
                                                                                                                                                 \draw [out =180,in=180,looseness=0.8,line width=1.7] (Geom) to node[near start,left=0.1cm,align=center]{}(Port);
                                                                                                                                                 \draw [-,line width=1.7] (Phys) to node[midway,above=0cm,align=center]{}(Math);
                                                                                                                                                 \draw [-,line width=1.7] (Chem) to node[near end,right=0.5cm,align=center]{}(Math);
                                                                                                                                                 \draw [-,line width=1.7] (Bio) to node[midway,above=0cm,align=center]{} (Chem);
                                                                                                                                                 \draw [-,line width=1.7] (Bio) to  node[midway,right=0.5cm,align=center]{} (Geo);
                                                                                                                                                 \draw[->] (Bio) to (RBio);
                                                                                                                                                 \draw[->] (Geo) to (RGeo);
                                                                                                                                                 \draw[->] (Chem) to (RChem);
                                                                                                                                                 \draw[->] (Math) to (RMath);
                                                                                                                                                 \draw[->] (Port) to (RPort);
                                                                                                                                                 \draw[->] (Geom) to (RGeom);
                                                                                                                                                 \draw[->] (Phys) to (RPhys);
                                                                                                                                                 \draw[->] (His) to (RHis);
                                                                                                                                                 \node[draw,inner xsep=5em,fit=(Port) (His) (Geo) (Chem) (Bio) (Math) (Phys) (Geom)] {};
 \end{tikzpicture}
 }
\end{center}
\caption{Independence graph  (central rectangle) representing the estimated graphical model describing the covariance structure of the individual random components $V_i^{[1]},\ldots,V_i^{[8]}$ for an  arbitrarily chosen individual  ($i=1, \ldots, n$, suppressing the index $i$ in the graph), for the students who did not receive bonus. 
}
\label{fig:AppNobonus}
\end{figure}
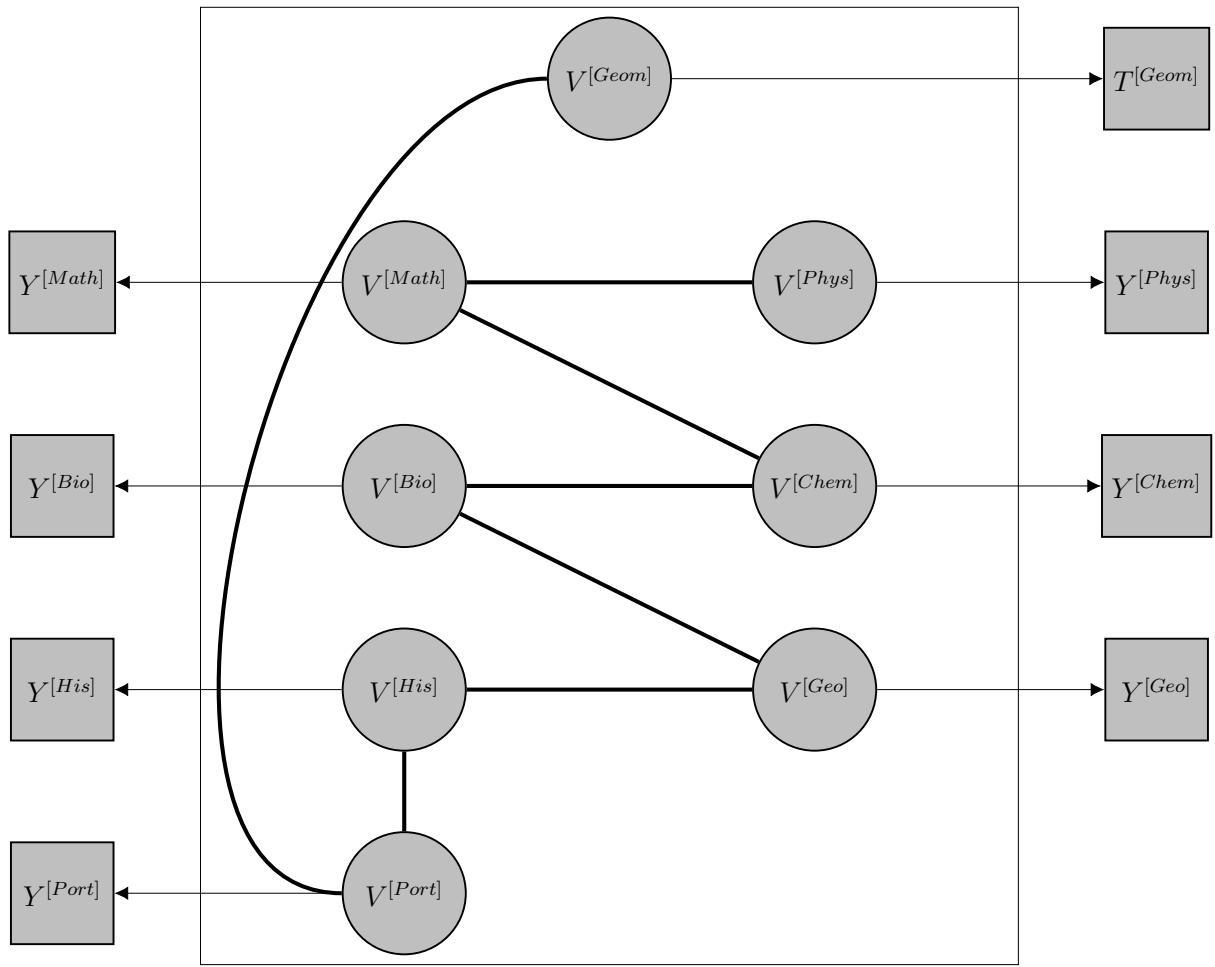

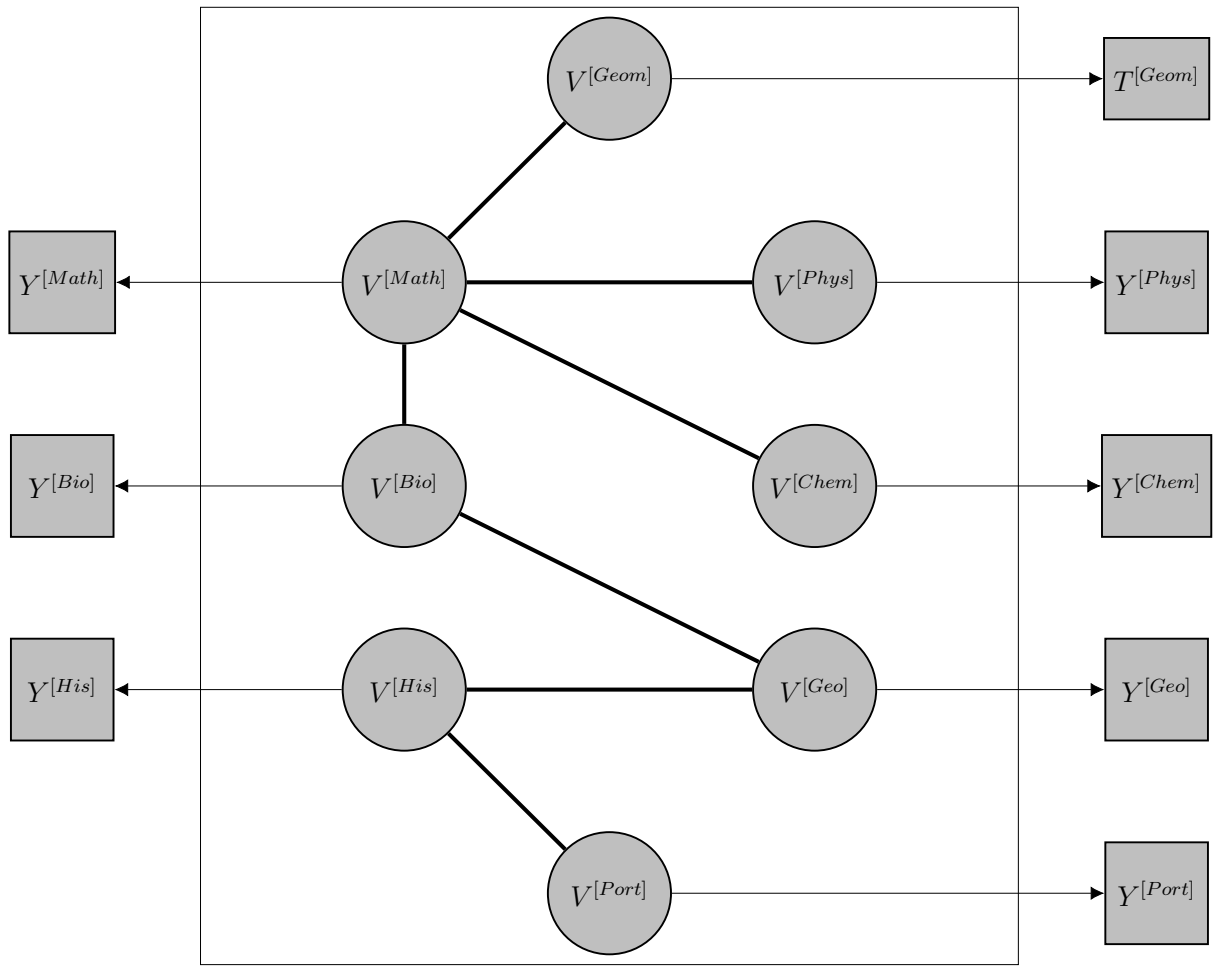
\begin{figure}
\begin{center}\scalebox{0.9}{

 \begin{tikzpicture}[
                                                                                                                                                 squarednode/.style={rectangle, draw=black, fill=lightgray, thick},
                                                                                                                                                 roundnode/.style={circle, draw=black, fill=lightgray, thick},
                                                                                                                                                 ]
                                                                                                                                                 \node [style=roundnode,minimum size=1.8cm] (His) at (-3, 3) {$V^{[His]}$};
                                                                                                                                                 \node [style=roundnode,minimum size=1.8cm] (Chem) at (3, 6) {$V^{[Chem]}$};
                                                                                                                                                 \node [style=roundnode,minimum size=1.8cm] (Port) at (0, 0) {$V^{[Port]}$};
                                                                                                                                                 \node [style=roundnode,minimum size=1.8cm] (Geo) at (3, 3) {$V^{[Geo]}$};
                                                                                                                                                 \node [style=roundnode,minimum size=1.8cm] (Math) at (-3,9) {$V^{[Math]}$};
                                                                                                                                                 \node [style=roundnode,minimum size=1.8cm] (Bio) at (-3, 6) {$V^{[Bio]}$};
                                                                                                                                                 \node [style=roundnode,minimum size=1.8cm] (Phys) at (3, 9) {$V^{[Phys]}$};
                                                                                                                                                 \node [style=roundnode,minimum size=1.8cm] (Geom) at (0, 12) {$V^{[Geom]}$};
                                                                                                                                                 \node [style=squarednode,minimum size=1.5cm] (RHis) at (-8, 3) {$Y^{[His]}$};
                                                                                                                                                 \node [style=squarednode,minimum size=1.5cm] (RChem) at (8, 6) {$Y^{[Chem]}$};
                                                                                                                                                 \node [style=squarednode,minimum size=1.5cm] (RPort) at (8, 0) {$Y^{[Port]}$};
                                                                                                                                                 \node [style=squarednode,minimum size=1.5cm] (RGeo) at (8, 3) {$Y^{[Geo]}$};
                                                                                                                                                 \node [style=squarednode,minimum size=1.5cm] (RMath) at (-8,9) {$Y^{[Math]}$};
                                                                                                                                                 \node [style=squarednode,minimum size=1.5cm] (RBio) at (-8, 6) {$Y^{[Bio]}$};
                                                                                                                                                 \node [style=squarednode,minimum size=1.5cm] (RPhys) at (8, 9) {$Y^{[Phys]}$};
                                                                                                                                                 \node [style=squarednode,minimum size=1.2cm] (RGeom) at (8, 12) {$T^{[Geom]}$};
                                                                                                                                                 \draw [-,line width=1.7] (His) to node[midway,above=0cm,align=center]{} (Geo);
                                                                                                                                                 \draw [-,line width=1.7] (His) to node[midway,left=0.1cm,align=center]{}(Port);
                                                                                                                   \draw [-,line width=1.7] (Geom) to node[midway,left=0.2cm,align=center]{}(Math);
                                                                                                                                                 \draw [-,line width=1.7] (Phys) to node[midway,above=0cm,align=center]{}(Math);
                                                                                                                                                 \draw [-,line width=1.7] (Bio) to node[midway,left=0cm,align=center]{}(Math);
                                                                                                                                                 \draw [-,line width=1.7] (Chem) to node[midway,right=0.5cm,align=center]{}(Math);
                                                                                                                                                 \draw [-,line width=1.7] (Bio) to  node[midway,right=0.5cm,align=center]{} (Geo);
                                                                                                                                                 \draw[->] (Bio) to (RBio);
                                                                                                                                                 \draw[->] (Chem) to (RChem);
                                                                                                                                                 \draw[->] (Geo) to (RGeo);
                                                                                                                                                 \draw[->] (Math) to (RMath);
                                                                                                                                                 \draw[->] (Phys) to (RPhys);
                                                                                                                                                 \draw[->] (Port) to (RPort);
                                                                                                                                                 \draw[->] (Geom) to (RGeom);
                                                                                                                                                 \draw[->] (His) to (RHis);
                                                                                                                                                 \node[draw,inner xsep=5em,fit=(Port) (His) (Geo) (Chem) (Bio) (Math) (Phys) (Geom)] {};
\end{tikzpicture}
 }
\end{center}
\caption{Independence graph  (central rectangle) representing the estimated graphical model describing the covariance structure of the individual random components $V_i^{[1]},\ldots,V_i^{[8]}$ for an  arbitrarily chosen individual  ($i=1, \ldots, n$, suppressing the index $i$ in the graph), for the students who received bonus. 
}
\label{fig:AppBonus}
\end{figure}

\end{document}